\documentclass[times,12pt]{article}
\usepackage[margin=1in]{geometry}
\usepackage[hidelinks]{hyperref}
\usepackage{lineno}
\usepackage{amsmath, amssymb}
\usepackage{hyperref}
\usepackage{graphicx}
\usepackage[round, comma]{natbib}
\usepackage{psfrag}
\usepackage[lofdepth,lotdepth]{subfig}
\usepackage{mathtools}
\usepackage{multirow, authblk}
\usepackage{multicol}
\usepackage{bm}
\usepackage{threeparttable, tablefootnote}
\usepackage{physics}
\usepackage{xcolor}
\usepackage{soul}
\extrafloats{100}
\usepackage{subfig}
\usepackage{chngcntr}
\usepackage{tabularx}
\newcommand{\GG}[1]{}
\hypersetup{
    colorlinks,
    linkcolor={red!80!black},
    citecolor={blue!50!black},
    urlcolor={blue!80!black}
}

\begin{document}

\begin{center}
    \textbf{Electron Impact Ionization in the Icy Galilean Satellites' Atmospheres} \\
\end{center}

\begin{center}
    Shane R. Carberry Mogan$^{1,*}$, Robert E. Johnson$^{2,3}$, Audrey Vorburger$^{4}$, Lorenz Roth$^{5}$ \\
\end{center}

\begin{center}
    $^1$Space Sciences Laboratory, University of California, Berkeley, Berkeley, USA; $^2$University of Virginia, Charlottesville, USA; $^3$New York University, New York, USA; $^4$University of Bern, Bern, Switzerland; $^5$KTH Royal Institute of Technology, Stockholm, Sweden; $^*$Corresponding author: Shane R. Carberry Mogan (CarberryMogan@Berkeley.edu)
\end{center}

\begin{abstract}

Electron impact ionization is critical in producing the ionospheres on many planetary bodies and, as discussed here, is critical for interpreting spacecraft and telescopic observations of the tenuous atmospheres of the icy Galilean satellites of Jupiter (Europa, Ganymede, and Callisto), which form an interesting planetary system. Fortunately, laboratory measurements, extrapolated by theoretical models, were developed and published over a number of years by K. Becker and colleagues (see \citealt{deutsch2009}) to provide accurate electron impact ionization cross sections for atoms and molecules, which are crucial to correctly interpret these measurements. Because of their relevance for the Jovian icy satellites we provide useful fits to the complex, semi-empirical Deutsch–M\"{a}rk formula for energy-dependent electron impact ionization cross-sections of gas-phase water products (i.e., H$_2$O, H$_2$, O$_2$, H, O). These are then used with measurements of the thermal plasma in the Jovian magnetosphere to produce ionization rates for comparison with solar photo-ionization rates at the icy Galilean satellites.

\end{abstract}

\section{Introduction}

Since Galileo Galilei's revolutionary discovery that Jupiter, the largest planet in the solar system, has four large planetary bodies revolving around it -- the ``Galilean'' satellites: Io, Europa, Ganymede, and Callisto -- our fascination with this planetary system has only grown with the advancement of observational technologies. Several spacecraft have been sent directly to or have at least passed by the Jovian system. In the 1970s, \textit{Pioneer 10}~\&~\textit{11} and \textit{Voyager 1}~\&~\textit{2} utilized Jupiter's gravity to enhance their trajectories and observed the giant planet and its moons up-close. From 1995--2003 \textit{Galileo} orbited Jupiter, and made several close encounters with the namesake moons. At the time of this writing, the Juno spacecraft is currently orbiting Jupiter and has recently made close flybys of Ganymede and Europa with 2 flybys of Io forthcoming. In addition to these in-situ observations, the Hubble Space Telescope (HST), which has been situated in Earth's orbit since 1990, has been used to observe and contribute new information to our understanding of this system. To better understand the Jovian system, the Galilean satellites, and their interconnected dynamics, as well as address certain prevailing mysteries, three forthcoming missions, ESA's JUpiter ICy moons Explorer, NASA's Europa Clipper, and CNSA's Gan De, will send spacecraft back to the Jovian system. 

Our focus here is on the \textit{icy} Galilean satellites (see Table \ref{tab:phys_param}) -- Europa, Ganymede, and Callisto -- and their tenuous atmospheres for which the dominant constituents are water products: H$_2$O, O$_2$, H$_2$, H, and O. Because these objects orbit Jupiter within its giant magnetic field, they are exposed to an ambient plasma. Interactions between this plasma and the icy Galilean satellites' atmospheres to a large extent determine the nature of the latter. One key aspect of this interaction is the role of the plasma electrons in dissociating, ionizing, and exciting the gas-phase water products via impacts, which can produce observable emission features (e.g., \citealt{hall1998, feldman2000, roth2017b, roth2021a, roth2021b}). In order to calculate the ionization rates to be used in future simulations of these atmospheres, we use the extensive laboratory data of electron impact ionization cross-sections accumulated and summarized by K. Becker and his colleagues over a number of years as reviewed in \cite{deutsch2009} and references herein. Since the various laboratory measurements can exhibit differences and typically cover only a limited range of energies, the group developed a fitting procedure also presented in those papers. In this method, the Deutsch–M\"{a}rk (DM) formula, atomic orbital basis sets are used in expressions to fit and extrapolate the measurements, as well as to generate cross-sections when measurements are unavailable. In this way, they created a large number of energy-dependent electron impact ionization cross-section distributions. Although the DM formula provides a broad range of useful data, as well as allows the calculation of results for molecular targets to be determined from the constituent atomic orbitals, it is not easily implemented. For example, in detailed molecular kinetic simulations much simpler calculations are more often made, such as those recently implemented in \cite{carberrymogan2022} for describing electron impacts in Callisto's atmosphere. More readily useful expressions are needed. Therefore, here we present much more simplified fits to the results obtained via the DM formula, which we then use with electron density and temperature data at the icy Galilean satellites to calculate ionization rates in their atmospheres. These rates are in turn needed to help interpret past, present, and future spacecraft and telescopic observations of these topical planetary bodies soon to be visited by several new spacecraft.

Before we discuss the derivation of the electron impact ionization cross-section fits (Section \ref{method}) and present the corresponding ionization rates for each species considered (Section \ref{results}), we first review the local space environment of the icy Galilean satellites, particularly the Jovian magnetospheric plasma in which they are embedded, as well as the observations of water products in their atmospheres.

\section{Background} \label{background}

The strong dynamo generated within the interior of Jupiter produces its magnetic field, which has a rotation period of $\sim$10 h. Jupiter's magnetosphere, i.e., the environment controlled by the planet's magnetic field, is filled with charged particles. While the volcanic moon Io is the main source, materials from the icy Galilean satellites' surfaces are also a source of charged particles (either directly ionized or lost as neutrals and ionized later in the magnetosphere) \citep{johnson2004,kivelson2004}, which are picked-up, accelerated, and then become ``trapped'' by this rapidly rotating field. The combination of this fast rotating magnetic field and the large number of charged particles trapped within it gives rise to the plasma detected in Jupiter's immense magnetosphere, which can extend as far as 45--100~$R_J$ from the planet, where $R_J = 71,492$~km is the radius of Jupiter. The Jovian magnetosphere is typically divided up into three regions: the inner ($<$10~$R_J$), the middle (10--40~$R_J$), and the outer ($>$40~$R_J$) regions \citep{khurana2004}; Europa resides within the inner region, while Ganymede and Callisto reside in the middle region. The Jovian magnetospheric plasma is typically described as being comprised of two populations: a ``cold'' thermal plasma with energies $<$~1~keV and a ``hot'' energetic plasma with energies $\geq$~1~keV \citep{krupp04,bagenal11}. Both populations are composed of electrons, as well as H$^+$, O$^{n+}$, and S$^{n+}$ ions \citep{Bagenal1981, Broadfoot1981,Nerney2017}. The sulfur and oxygen ions, both of which are of various charge states, primarily originate from the volcanic moon Io \citep{bagenal2020}. On the other hand, the hydrogen ions originate from two different sources depending on the magnetospheric region: in the inner region, they mainly originate in Jupiter’s atmosphere; and in the middle and outer magnetosphere, the solar wind can gain access to the magnetosphere, becoming the main provider of hydrogen ions thereafter. An additional plasma contribution (electrons, H$^+$, O$^{n+}$) is associated with sputtering of the icy satellite’s surfaces \citep{Cooper2001, johnson2004, szalay2022}. Here we focus on the thermal electrons as these particles are the most relevant to ionization. Physical parameters for thermal electrons at the orbits of Europa, Ganymede, and Callisto based on \textit{Voyager} and \textit{Galileo} data are listed in Table \ref{tab:phys_param}.

\setlength{\tabcolsep}{10pt}
\renewcommand{\arraystretch}{1.5}
\begin{table}[h!]
\centering
\caption{Physical Parameters of Europa, Ganymede, and Callisto, as well as of the Jovian magnetosphere at their orbits}
\begin{tabular}{|l|c|c|c|c|}
\hline
\textbf{Parameters [units]} & \textbf{Europa} & \textbf{Ganymede} & \textbf{Callisto} \\
\hline
Radius [km] & 1,561 & 2,631 & 2,410\\
\hline
Mass [($\times$10$^{23}$) kg] & 0.48 & 1.5 & 1.1\\
\hline 
Distance from Jupiter [$R_J$] & 9.38 & 15.0 & 26.3\\
\hline
Orbital Velocity [km s$^{-1}$] & 13.7 & 10.9 & 8.20 \\
\hline
Thermal Electron Density [cm$^{-3}$] & 18--290 $^{a, b}$ \tnote{a,b} & 1--10 $^{a, c}$ & 0.01--1.1 $^{a, c}$ \\
\hline
Thermal Electron Temperature [eV] & 10--30 $^{b}$ & 100 $^{c}$ \tnote{c} & 100 $^{c}$ \\
\hline
Thermal Electron Flux $^d$ \tnote{d} [cm$^{-2}$ s$^{-1}$] & (0.4--1.1)$\times$10$^{11}$ $^e$ \tnote{e} & (0.7--6.7)$\times$10$^9$ & (0.08--7.4)$\times$10$^8$ \\
\hline
Plasma Azimuthal Velocity [km s$^{-1}$] & 90 $^a$ & 150 $^a$ & 200 $^a$ \\
\hline
Relative Plasma Velocity $^f$ \tnote{f} [km s$^{-1}$] & 76 $^a$ & 139 $^a$ & 192 $^a$ \\
\hline
\end{tabular}
\label{tab:phys_param}
    \begin{tablenotes}\footnotesize
        \item[a]$^a$ Values taken from \cite{kivelson2004} and references therein.
        \item[b]$^b$ Values taken from \cite{bagenal2015}.
        \item[c]$^c$ Values taken from \cite{neubauer1998} and references therein.
        \item[d]$^d$ Thermal electron flux, $\phi_e = n_e v_e$, where $n_e$ is the thermal electron density, $v_e = \sqrt{8 k_B T_e / \pi / m_e}$ is the mean Maxwellian speed of the electrons, with $k_B T_e$ the thermal electron temperature, $k_B$ the Boltzmann constant, and $m_e$ the mass of an electron.
        \item[e]$^e$ $\phi_e$ calculated according to the ``low/hot'' ($k_B T_e = 30$~eV, $n_e = 63$~cm$^{-3}$) and ``high/cold'' ($k_B T_e$~=~10~eV, $n_e = 290$~cm$^{-3}$) plasma components from Table 5 in \cite{bagenal2015}.
        \item[g]$^g$ Relative speed between plasma azimuthal velocity and the satellites' orbital speeds.
    \end{tablenotes}
\end{table}

As a result of the large relative velocities between the azimuthal velocity of the Jovian magnetosphere and the icy Galilean satellites' orbital velocities (Table \ref{tab:phys_param}), the satellites are continuously overtaken and bombarded by the magnetospheric plasma. The spatial distribution of this bombardment is determined by the flow rate of the plasma particles past the satellite as well as their thermal velocities relative to the local magnetic field lines \citep{johnson2004}. However, intrinsic or induced electric and magnetic fields as well as the interactions with the tenuous atmospheres and ionospheres at these satellites can radically modify the local fluxes of impinging particles at Ganymede (e.g., from \citealt{paranicas2022}) and at Callisto (e.g., from \citealt{strobel2002}). The cyclotron radii or gyro-radii of these charged particles depend on their mass, speed, and charge, as well as the local magnetic field strength. Due to their small mass, electrons primarily have gyro-radii much smaller than the satellite radius, and thus preferentially impact the satellites' trailing hemispheres and poles as they move up and down the rotating field lines \citep{johnson2004}.

Following the \textit{Pioneer} discovery of intense plasma trapped in the Jovian magnetic field \citep{smith1974, wolfe1974, trainor1974, frank1976} a series of experiments were carried out to measure the effect this could have on the icy surfaces of Europa, Ganymede, and Callisto \citep{brown1978, lanzerotti1978}. These experiments showed that the ejection of water molecules from low-temperature ices by incident energetic particles, a process referred to as ``sputtering'', is dominated by electronic excitations and ionizations produced in the ice (``electronic'' sputtering), rather than by ``knock-on'' collisions of the ions with water molecules (``nuclear'' sputtering), the hitherto typically studied sputtering process. Subsequent experiments led to the discovery that additional molecular species can form in and be released from the ice, namely H$_2$ and O$_2$, in a process referred to as ``radiolysis'' \citep{brown1982, boring1983, reimann1984, brown1984}, which occurs as bonds in H$_2$O molecules are broken by the electronic energy deposited by the impinging charged particles and the fragmented molecules recombine. Moreover, the number of radiolytic products ejected from the icy surface per each incident charged particle (i.e., the ``sputter yield'') was shown to display a strong temperature dependence.

These discoveries had immense implications for the icy Galilean satellites: magnetospheric plasma-induced sputtering could erode their surfaces, and the ejected atoms and molecules could migrate significant distances as well as escape the local gravitational environment of its host satellite or form gravitationally bound atmospheres \citep{johnson1990}. Indeed Europa was predicted to have a tenuous, predominantly O$_2$ atmosphere due to the radiolytic decomposition of its icy surface by the incident Jovian plasma particles \citep{johnson1982, johnson1983, johnson1990}, which has been borne out by extensive HST observations (e.g., \citealt{roth2016}, and references therein). Tenuous O$_2$ atmospheres produced via similar processes have also been detected at Ganymede and Callisto: following the HST detection of O emissions indicative of an O$_2$ atmosphere at Europa \citep{hall1995}, airglow emissions were detected by HST in Ganymede's O$_2$ atmosphere \citep{hall1998} as well as in Europa's atmosphere thereby confirming the earlier observation; O emissions were detected by HST in Callisto's atmosphere \citep{cunningham2015}, which were suggested to be induced by photoelectron impacts in a near-surface, O$_2$-dominated atmosphere; and recently atomic O emissions have been detected at Ganymede \citep{roth2021b} indicative of being produced via dissociative excitations of O$_2$ (and H$_2$O).

Although H$_2$ can more readily escape from the atmospheres of these bodies than can the concomitant radiolytically produced O$_2$, a steady-state H$_2$ atmospheric component can also form (e.g., \citealt{carberrymogan2022}). Atomic H, the dissociated product of H$_2$, has also been detected at Callisto (e.g., \citealt{roth2017b, carberrymogan2022}). Moreover, \cite{carberrymogan2022} and \cite{roth2023} recently suggested that the H detected in the extended atmospheres of Europa \citep{roth2017b, roth2023} and Ganymede \citep{barth1997, feldman2000, alday2017, roth2023} are also indicative of an H$_2$-source. Although H$_2$ is able to escape from these satellites' atmospheres, it does not escape from the Jovian system; and since its lifetime is longer than the satellites' orbital periods (e.g., \cite{smyth2006, leblanc2017, carberrymogan2022}), a detectable toroidal cloud of neutral H$_2$ co-rotating with the bodies can form (e.g., \citealt{szalay2022}).

The sputtering and radiolytic sources of the icy Galilean satellites' atmospheres compete with other sources, such as sublimation of water ice and the subsequent photochemistry of the newly formed water vapor (e.g., \citealt{yung1977, kumar1982}). However, with increasing distance from Jupiter, the plasma density and, as a result, the corresponding atmospheric source decreases \citep{johnson2004}. For example, although gas-phase H$_2$O has not been directly observed at Callisto, the outermost Galilean satellite, observed geomorphological features have been interpreted to be caused by sublimation of the surface ice rather than by sputtering \citep{spencer1984, spencer1987, moore1999}. Further, whereas sublimation has been suggested to be the primary source of Ganymede's H$_2$O atmosphere \citep{roth2021b, vorburger2022}, sputtering has been suggested to be a primary source of Europa's H$_2$O atmosphere \citep{addison2021}. 

Below we focus on deriving thermal electron impact ionization rates in these icy satellites' atmospheres, which are needed to help understand their evolution.

\section{Method} \label{method}

The Deutsch-M\"{a}rk (DM) formula was developed to employ and extrapolate laboratory data to allow users to estimate reasonably accurate electron impact ionization cross-sections, $\sigma(E)$, over a large range of energies, $E$. The ``modified'' DM formula (Appendix \ref{app:DM}) from \cite{deutsch2004}, hereafter referred to as ``DM2004,'' is a revised version of the original formula developed by \cite{deutsch1987} and used to calculate cross-sections up to energies $\lesssim$ keV. As discussed by \cite{deutsch2000}, hereafter referred to as ``DM2000,'' and references therein, $\sigma(E)$ for atoms (e.g., H and O) can be converted to $\sigma(E)$ for molecules composed of those atoms (e.g., H$_2$O, O$_2$, and H$_2$).

Here we present more easily usable fits to these complex models for species of interest to the planetary science community, and of particular interest at icy satellites. An exponentially modified Gaussian distribution is used to compute a non-linear least-squares fit to $\sigma(E)$ derived for H, O, H$_2$, O$_2$, and H$_2$O using DM2000 and DM2004. The resulting equation is as follows (in units of $\times$10$^{-16}$ cm$^2$):

\begin{equation}
    \sigma(E) = \frac{\alpha}{2 \delta} \exp \left( \frac{\gamma^2}{2 \delta^2} + \frac{\beta - \log_{10}(E)}{\delta} \right) \left( \erf \left( \frac{\log_{10}(E) - \beta}{\sqrt{2} \gamma} - \frac{\gamma}{\sqrt{2} \delta} \right) + \frac{\delta}{\abs{\delta}} \right) + \epsilon,
    \label{fit_eq}
\end{equation}

\noindent where the coefficients $\alpha$, $\beta$, $\gamma$, $\delta$, and $\epsilon$ in Eq. \ref{fit_eq} for H, O, H$_2$, O$_2$, and H$_2$O are listed in Table \ref{tab:fit_eq_coeff}.

\begin{table}[h!]
    \centering
    \caption{Coefficients in Eq. \ref{fit_eq} for H, O, H$_2$, O$_2$, H$_2$O}
    \begin{tabular}{|c|c|c|c|c|c|}
        \hline
        \textbf{Species} & $\alpha$ & $\beta$ & $\gamma$ & $\delta$ & $\epsilon$ \\
        \hline
        H & 0.951653 & 1.40862 & 0.271538 & 0.804953 & -0.0397646 \\
        \hline
        O & 1.91899 & 1.72847 & 0.333420 & 0.864596 & -0.121378 \\
        \hline
        H$_2$ & 2.27256 & 1.39600 & 0.277991 & 1.08255 & -0.278929 \\
        \hline
        O$_2$ & 3.20403 & 1.63262 & 0.241759 & 0.799914 & -0.0876336 \\
        \hline
        H$_2$O & 4.41745 & 1.48743 & 0.291951 & 1.01222 & -0.527864 \\
        \hline
    \end{tabular}
    \label{tab:fit_eq_coeff}
\end{table}

\section{Results} \label{results}

\begin{figure}[h!]
    \centering
    \includegraphics[width=\textwidth]{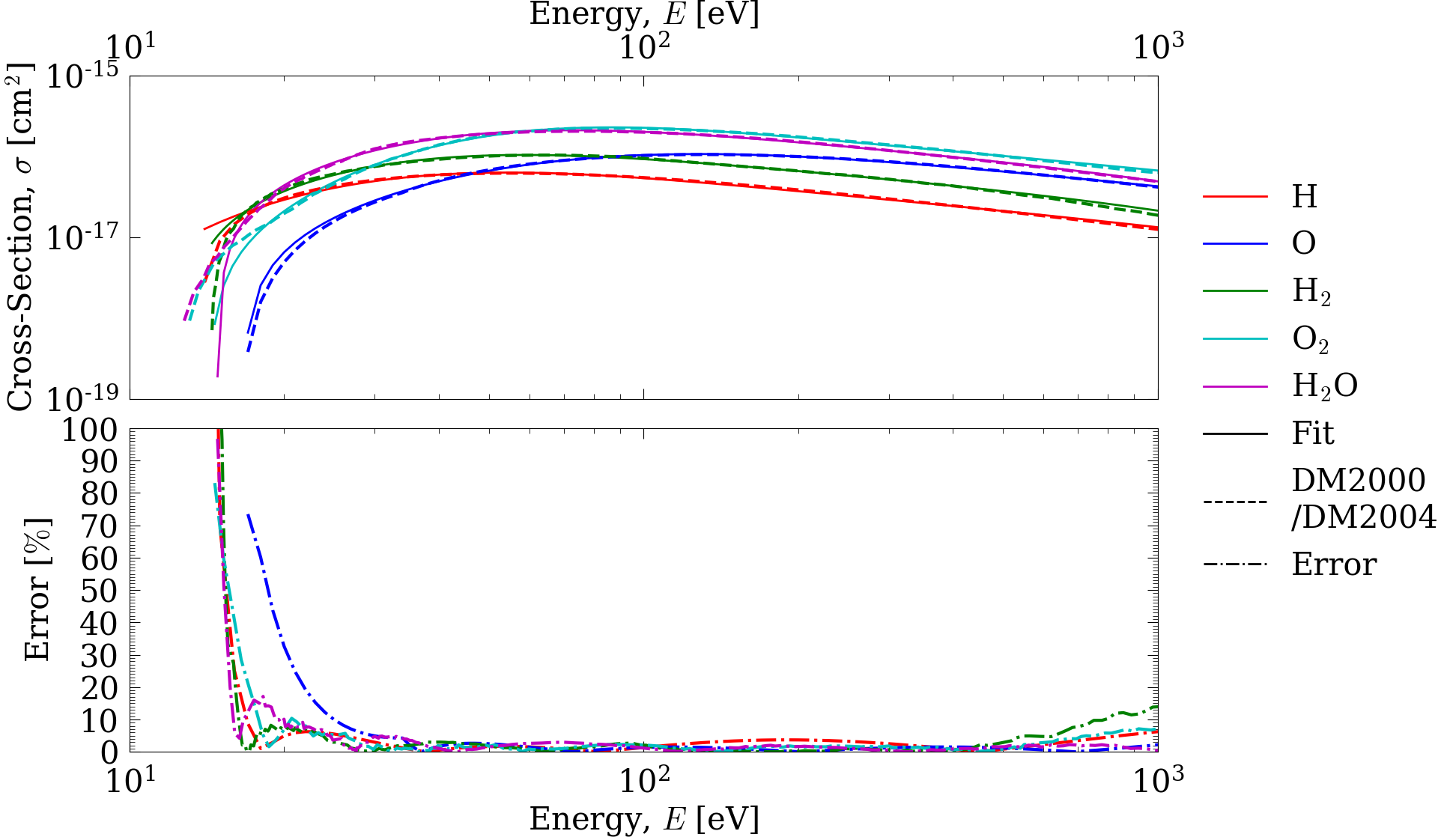}
    \caption{Top panel: Electron impact ionization cross-section, $\sigma(E)$, for H (red lines), O (blue lines), H$_2$ (green lines), O$_2$ (cyan lines), and H$_2$O (magenta lines) as a function of incident electron energy, $E$ (x-axis). The dashed colored lines are from \cite{deutsch2000} (``DM2000'') for H$_2$, O$_2$, and H$_2$O and from \cite{deutsch2004} (``DM2004'') for H and O; and the solid colored lines are the corresponding fits (``Fit'') calculated via Eq. \ref{fit_eq} with the coefficients from Table \ref{tab:fit_eq_coeff}. Note the values for $\sigma(E)$ begin at the ionization energies of the species considered (e.g., Tables~\ref{tab:H_DM_eq1}--\ref{tab:O_DM_eq1} in Appendix~\ref{app:DM}), hence the blank spaces below these energies. Bottom panel: The dash-dotted colored lines represent the difference between $\sigma(E)$ calculated via DM2000/DM2004 and via Eq. \ref{fit_eq}, $\frac{|\mathrm{fit}-\mathrm{DM}2000/2004|}{\mathrm{DM}2000/2004}\times$100 (``Error'').}
    \label{fig:fit_vs_data}
\end{figure}

Electron impact ionization cross-sections, $\sigma(E)$, for H, O, H$_2$, O$_2$, and H$_2$O as derived by DM2000 and DM2004, as well as the corresponding fits calculated via Eq. \ref{fit_eq} with the coefficients from Table \ref{tab:fit_eq_coeff}, are illustrated in Figure \ref{fig:fit_vs_data}. These fits, of course, account for the ionization threshold energies occurring around 10 and 20 eV for the species considered (e.g., see Tables \ref{tab:H_DM_eq1}--\ref{tab:O_DM_eq1} in Appendix \ref{app:DM}). By about $\sim$20 eV, the difference in $\sigma(E)$ derived by DM2000/DM2004 and by the corresponding fits for all of the species considered fall below 10$\%$, except for that of O, which drops below 10$\%$ between 20 and 30 eV. From these lower bounds up to 1 keV, the difference remains below 10$\%$ for all species considered except for H$_2$, for which it exceeds 10$\%$ by $\sim$800 eV but is only $\sim$14$\%$ by 1 keV. Thus, between $\sim$20~eV--1~keV (i.e., between the ionization energy of the species considered, (e.g., Tables~\ref{tab:H_DM_eq1}--\ref{tab:O_DM_eq1} in Appendix~\ref{app:DM}), and the maximum energy of the thermal electrons in the Jovian magnetosphere at the icy Galilean satellites, Table \ref{tab:phys_param}), the fits provided here can determine electron impact ionization cross-sections within $\sim$10$\%$ accuracy of those determined via the more complex DM models, which have been extensively tested, modified, and improved over the years, demonstrating agreement with experimental data better than $\sim$20--35$\%$ (\citealt{deutsch2009} and references therein).

With the thermal electron fluxes and temperatures listed in Table \ref{tab:phys_param}, we use the fits for electron impact ionization cross-sections presented in Fig. \ref{fig:fit_vs_data} to determine the corresponding ionization rates in the water product atmospheres of Europa, Ganymede, and Callisto, which are presented in Table \ref{tab:ele_imp_rates}. The difference in $\sigma(E)$ derived by DM2004 and by the corresponding fit for O electron impact ionization cross-section at 20 eV is $\sim$32$\%$, making the latter (Eq.~\ref{fit_eq}) in the region of Europa's orbit not as accurate as those of the other species, for which the difference is always $<$ 10$\%$ (Table \ref{tab:ele_imp_rates}). However, by 30 eV the difference for all species drops to $<$ 5$\%$. The differences for electron impact ionization rates at incident electron energies of 100 eV at Ganymede and Callisto are $<$ 3$\%$ for all species. We compare these electron impact ionization rates to the analogous photoionization rates derived according to solar activity \citep{huebner2015} in Table \ref{tab:reactions} in Appendix~\ref{app:photo}. Since the electron temperature for the ``high/cold'' plasma component at Europa (Table \ref{tab:phys_param}) is less than the ionization energies of the atoms and molecules considered (e.g., Tables \ref{tab:H_DM_eq1} and \ref{tab:O_DM_eq1} in Appendix \ref{app:DM}), we show ionization rates over a temperature range of 20--30 eV (``medium'' to ``low/hot'' plasma components from \citealt{bagenal2015}). At Europa and Ganymede, electron impact ionization rates are greater than the photoionization rates for all species. On the other hand, at Callisto, where there is the most uncertainty in electron densities (see e.g., Table \ref{tab:phys_param}), the upper bound electron impact ionization rates are greater than the upper bound photoionization rates for all species, but the lower bound electron impact ionization rates are less than the lower bound photoionization rates. We note, however, that the electron impact ionization rates relate to the upstream plasma properties, and the effective electron impact ionization strongly depends on the details of the interaction of the plasma flow with the moons' atmospheres and ionospheres, which cool as well as divert the plasma around the moons (e.g., \citealt{saur1998, saur2004, rubin2015, dols2016}).

\begin{table}[h!]
    \centering
    \caption{Electron impact ionization cross-sections and rates in the water product atmospheres of Europa (E), Ganymede (G), and Callisto (C)}
    \begin{tabular}{|c|c|c|c|c|}
    \hline
    {\textbf{Species}} & \textbf{Satellite} & \textbf{Cross-Section}$^a$ [cm$^{-2}$] & \tnote{a} \textbf{Rate}$^b$ \tnote{b} [s$^{-1}$] & \textbf{Error}$^c$ \tnote{c} [$\%$] \\
    \hline
    \multirow{3}{*}{H} & E & (3.07--5.11)$\times$10$^{-17}$ & (2.57--5.57)$\times$10$^{-6}$ & 2.96--4.93 \\ \cline{2-5}
    & G & \multirow{2}{*}{5.48$\times$10$^{-17}$} & (0.384--3.67)$\times$10$^{-7}$ & \multirow{2}{*}{1.34} \\ \cline{2-2}\cline{4-4}
    & C & & 4.38$\times$10$^{-10}$ -- 4.06$\times$10$^{-8}$ & \\
    \hline
    \multirow{3}{*}{O} & E & (0.495--2.72)$\times$10$^{-17}$ & (0.414--2.97)$\times$10$^{-6}$ & 4.92--32.6 \\ \cline{2-5}
    & G & \multirow{2}{*}{1.03$\times$10$^{-16}$} & (0.719--6.88)$\times$10$^{-7}$ & \multirow{2}{*}{1.36} \\ \cline{2-2}\cline{4-4}
    & C & & 8.22$\times$10$^{-10}$ -- 7.60$\times$10$^{-8}$ & \\
    \hline
    \multirow{3}{*}{H$_2$} & E & (3.99--7.52)$\times$10$^{-17}$ & (3.34--8.20)$\times$10$^{-6}$ & 4.64--6.86 \\ \cline{2-5}
    & G & \multirow{2}{*}{9.48$\times$10$^{-17}$} & (0.664--6.35)$\times$10$^{-7}$ & \multirow{2}{*}{2.04} \\ \cline{2-2}\cline{4-4}
    & C & & 7.58$\times$10$^{-10}$ -- 7.01$\times$10$^{-8}$ & \\
    \hline
    \multirow{3}{*}{O$_2$} & E & (1.98--7.80)$\times$10$^{-17}$ & (1.66--8.50)$\times$10$^{-6}$ & 1.22--6.91 \\ \cline{2-5}
    & G & \multirow{2}{*}{2.22$\times$10$^{-16}$} & (0.155--1.49)$\times$10$^{-6}$ & \multirow{2}{*}{2.12}\\ \cline{2-2}\cline{4-4}
    & C & & 1.77$\times$10$^{-9}$ -- 1.64$\times$10$^{-7}$ &    \\
    \hline
    \multirow{3}{*}{H$_2$O} & E & (0.401--1.24)$\times$10$^{-16}$ & (0.336--1.36)$\times$10$^{-5}$ & 4.63--9.82 \\ \cline{2-5}
    & G & \multirow{2}{*}{2.00$\times$10$^{-16}$} & (0.140--1.34)$\times$10$^{-6}$ & \multirow{2}{*}{1.16}\\ \cline{2-2}\cline{4-4}
    & C & & 1.60$\times$10$^{-9}$ -- 1.47$\times$10$^{-7}$ &    \\
    \hline
    \end{tabular}
    \label{tab:ele_imp_rates}
    \begin{tablenotes}\footnotesize
        \item[a]$^a$ Electron impact ionization cross-sections are calculated via Eq.~\ref{fit_eq} at 20--30 eV for Europa and 100~eV for Ganymede and Callisto. Note we only consider temperatures $k_B T_e \ge 20$ eV at Europa because the minimum temperature, $k_B T_e = 10$ eV (Table~\ref{tab:phys_param}), is lower than the ionization energies of the species considered (e.g., Tables~\ref{tab:H_DM_eq1}--\ref{tab:O_DM_eq1} in Appendix~\ref{app:DM}).
        \item[b] $^b$ Electron impact ionization rates are calculated as the product of the electron impact ionization cross-sections and the range in thermal electron fluxes given in Table \ref{tab:phys_param}. Note, since the minimum temperature at Europa, $k_B T_e = 10$ eV, is lower than the ionization energies of the species considered, the lower bound thermal electron flux is calculated according to the ``medium'' plasma component ($k_B T_e = 20$ eV, $n_e = 158$ cm$^{-3}$) from \cite{bagenal2015}, Table 5 therein.
        \item[c]$^c$ The differences in the cross-sections derived by DM2000/DM2004 and by the corresponding fits, the ``errors,'' are interpolated from that illustrated in Fig. \ref{fig:fit_vs_data} at 20--30 eV for Europa (with the lower value calculated at 30 eV) and 100 eV for Ganymede and Callisto.
    \end{tablenotes}
\end{table}

\section{Conclusion} \label{conclusion}

The importance to the space physics community of data on atomic and molecular processes driven by an ambient plasma cannot be overstated. There are so many difficult but important observations whose interpretation is limited by the uncertainties in the atomic and molecular database or by the limited range of energies and species studied in the laboratory. Therefore, the combination of laboratory measurements with detailed, physically-based extrapolation procedures, as carried out by K. Becker and colleagues, will continue to be incredibly useful. Because the accurate DM formula used to develop useful electron impact cross sections over a large range of energies requires a considerable understanding of atomic physics, here we present more readily useful fits to their detailed analyses for use by the space physics community in order to prepare for the expected data from the forthcoming observations of plasma-atmosphere interactions at the icy Jovian satellites. These are used to show the relative importance of electron impact ionization in the icy Galilean satellites' atmospheres as compared to photo-ionization.

Finally, this study can be summarized as followed:

\begin{itemize}
    
    \item Fits to the Deutsch–M\"{a}rk formula for energy-dependent electron impact ionization cross-sections have been derived for H$_2$O, H$_2$, O$_2$, H, O from the species' minimum ionization energies up to 1 keV.
    
    \item These cross-sections are used in tandem with electron data at the orbits of Europa, Ganymede, and Callisto to determine the corresponding electron impact ionization rates in these bodies' water-product atmospheres.
    
    \item At Europa and Ganymede the electron impact ionization rates are shown to exceed the photoionization rates, whereas at Callisto, where the electron densities vary the most, likely a result of the moon being inside or outside of the Jovian plasma sheet, the electron impact ionization rates can be more or less than the photoionization rates.
    
\end{itemize}

\newpage

\appendix
\counterwithin{figure}{section}
\section*{Appendix}

\section{DM Formula} \label{app:DM}

\setcounter{table}{0}
\renewcommand*\thetable{\Alph{section}.\arabic{table}}

The ``modified'' DM formula \citep{deutsch2004} derives the total energy-dependent electron-impact ionization cross section, $\sigma(E)$, of an atom as:

\begin{equation}
    \sigma(E) = \sum_{n,l} \pi g_{n,l} r_{n,l}^2 \xi_{n,l} b_{n,l}^{(q)}(u) \left[ \ln (c_{n,l} u) / u \right].
    \label{DM_eq1}
\end{equation}

\noindent Here $r_{n,l}$ is the radius of maximum radial density of and $\xi_{n,l}$ is the number of electrons in the atomic subshell characterized by quantum numbers $n$ and $l$; $g_{n,l}$ is a weighting factor originally determined from a fitting procedure; $u = E / E_{n,l}$, where $E$ is the incident energy of the electrons and $E_{n,l}$ is the ionization energy in the ($n$, $l$) subshell; and $c_{n,l}$ is a constant determined from measured cross-sections for various values of $n$ and $l$. Tables \ref{tab:H_DM_eq1} and \ref{tab:O_DM_eq1} list values for the various terms in Eq. \ref{DM_eq1} for H and O atoms, respectively. The energy-dependent function $b_{n,l}^{(q)}(u) \left[ \ln (c_{n,l} u) / u \right]$ allows the DM formula to be applied up to keV-energy regimes, with $b_{n,l}^{(q)}(u)$ written as the following:

\begin{equation}
    b_{n,l}^{(q)}(u) = \frac{A_1 - A_2}{1 + (u / A_3)^p} + A_2,
    \label{DM_eq2}
\end{equation}

\noindent where $A_{1-3}$ and $p$ are constants determined from measured cross-sections for various values of $n$ and $l$, and the superscript $(q)$ refers to the number of electrons in the ($n$, $l$) subshell. Tables \ref{tab:H_DM_eq2} and \ref{tab:O_DM_eq2} list values for the various terms in the energy-dependent function $b_{n,l}^{(q)}(u) \left[ \ln (c_{n,l} u) / u \right]$ for H and O atoms, respectively. We refer the reader to the review by \cite{deutsch2009} for how $\sigma(E)$ of atoms are used to calculate $\sigma(E)$ of molecules composed of those atoms; i.e., how to estimate $\sigma(E)$ of H$_2$, O$_2$, and H$_2$O from $\sigma(E)$ of H and O. 

\newpage

\begin{table}[h!]
    \centering
    \caption{Various terms in Eq. \ref{DM_eq1} for electron impact ionization cross-section of H atoms}
    \begin{tabular}{|c|c|c|c|c|c|}
        \hline
        $n$ & $l$ & $\xi_{n,l}$ & $E_{n,l}$ $^a$ [eV] & $r_{n,l}$ $^a$\tnote{a} [($\times 10^{-11}$) m] & $g_{n,l}$ $^b$ \tnote{b} \\
        \hline
        1 & 0 & 1 & 13.6 & 5.29 & 2.81 \\
        \hline        
    \end{tabular}
    \label{tab:H_DM_eq1}
    \begin{tablenotes}\footnotesize
    \item[a]$^a$ $E_{n,l}$ and $r_{n,l}$ are taken from Tables 2 and 4 in \cite{desclaux1973}, respectively.
    \item[b]$^b$ $g_{n,l}$ is determined by dividing $E_{n,l}$ from the ``reduced weighting factor'' $g_{n,l} E_{n,l} = 38.20$ for 1$s^1$ in \cite{deutsch2000}.
    \end{tablenotes}
\end{table}

\begin{table}[h!]
    \centering
    \caption{Various terms in Eq. \ref{DM_eq1} for electron impact ionization cross-section of O atoms}
    \begin{tabular}{|c|c|c|c|c|c|}
        \hline
        $n$ & $l$ & $\xi_{n,l}$ & $E_{n,l}$ $^a$ [eV] & $r_{n,l}$ $^a$\tnote{a} [($\times 10^{-11}$) m] & $g_{n,l}$ $^b$ \tnote{b} \\
        \hline
        1 & 0 & 2 & 563 & 0.684 & 0.124 \\
        \hline
        2 & 0 & 2 & 34.1 & 4.63 & 0.587 \\
        \hline
        2 & 1 & 4 & 16.7 & 4.41 & 1.79 \\
        \hline        
    \end{tabular}
    \label{tab:O_DM_eq1}
    \begin{tablenotes}\footnotesize
    \item[a]$^a$ $E_{n,l}$ and $r_{n,l}$ are taken from Tables 2 and 4 in \cite{desclaux1973}, respectively.
    \item[b]$^b$ $g_{n,l}$ is determined by dividing $E_{n,l}$ from the ``reduced weighting factors'' $g_{n,l} E_{n,l}$ = 70.00, 20.00, and 30.00 for 1$s^2$, 2$s^2$, and 2$p^4$, respectively, in \cite{deutsch2000}.
    \end{tablenotes}
\end{table}

\begin{table}[h!]
    \centering
    \caption{Various terms in the energy-dependent function $b_{n,l}^{(q)}(u) \left[ \ln (c_{n,l} u) / u \right]$ (Eqs. \ref{DM_eq1}--\ref{DM_eq2}) for electron impact ionization cross-section of H atoms}
    \begin{tabular}{|c|c|c|c|c|c|c|c|}
        \hline
        $n$ & $l$ & $q$ & $c_{n,l}$ & $A_1$ & $A_2$ & $A_3$ & $p$ \\
        \hline
        1 & 0 & 1 & 1.00 & 0.31 & 0.87 & 2.32 & 1.95 \\
        \hline        
    \end{tabular}
    \label{tab:H_DM_eq2}
\end{table}

\begin{table}[h!]
    \centering
    \caption{Various terms in the energy-dependent function $b_{n,l}^{(q)}(u) \left[ \ln (c_{n,l} u) / u \right]$ (Eqs. \ref{DM_eq1}--\ref{DM_eq2}) for electron impact ionization cross-section of O atoms}
    \begin{tabular}{|c|c|c|c|c|c|c|c|}
        \hline
        $n$ & $l$ & $q$ & $c_{n,l}$ & $A_1$ & $A_2$ & $A_3$ & $p$ \\
        \hline
        1 & 0 & 2 & 1.01 & 0.23 & 0.86 & 3.67 & 2.08 \\
        \hline
        2 & 0 & 2 & 1.01 & 0.23 & 0.86 & 3.67 & 2.08 \\
        \hline
        2 & 1 & 4 & 1.02 & -0.15 & 1.17 & 4.05 & 1.31 \\
        \hline        
    \end{tabular}
    \label{tab:O_DM_eq2}
\end{table}

\newpage

\section{Photoionization Rates} \label{app:photo}

\setcounter{table}{0}

Table \ref{tab:reactions} lists the range of photoionization rates determined by \cite{huebner2015} for a ``quiet'' Sun (i.e., solar minimum) -- ``active'' Sun (i.e., solar maximum), which are then scaled to the average Jovian system's distance from the Sun, 5.2 AU, ignoring any possible absorption with depth into the atmosphere.

\begin{table}[h!]
    \centering
    \caption{Photoionization rates at 5.2 AU}
    \begin{tabular}{|c|c|}
        \hline
        {\textbf{Species}} & \textbf{Rate} [s$^{-1}$] \\
        \hline
        H       & (2.68--6.36)$\times$10$^{-9}$ \\
        \hline
        O       & (0.880--2.44)$\times$10$^{-8}$ \\
        \hline
        H$_2$   & (2.00--4.25)$\times$10$^{-9}$ \\
        \hline
        O$_2$   & (1.73--4.36)$\times$10$^{-8}$ \\
        \hline
        H$_2$O  & (1.22--3.06)$\times$10$^{-8}$ \\
        \hline        
    \end{tabular}
    \label{tab:reactions}
\end{table}

\section*{Acknowledgments}

S.R.C.M. acknowledges the support provided by NASA through the Solar System Workings grant 80NSSC21K0152, A.V. acknowledges the support provided by the Swiss National Science Foundation, and L.R. was supported by the Swedish National Space Agency through grant 2021-00153 and by the Swedish Research Council through grant 2017-04897. 

\section*{Author Contribution Statement}

All authors contributed equally to this work.

\section*{Data Availability Statement}

All data generated or analyzed during this study are included in this published article.

\newpage

\bibliography{References.bib}

\end{document}